\documentclass[12pt,letterpaper]{article}
\usepackage{amsmath,amsfonts}

\title{Spinor Parallel Propagator and Green's Function in Maximally
Symmetric Spaces}
\author{Wolfgang M\"uck\thanks{E-mail address: wmueck@sfu.ca}\\
\small \textit{Department of Physics, Simon Fraser University,
Burnaby, B.C., V5A 1S6 Canada}}
\date{December 8, 1999}

\providecommand{\slashD}{D\mspace{-12mu}\slash\mspace{3mu}}
\providecommand{\slashdel}{\partial\mspace{-9mu}\slash}
\providecommand{\sgn}{\mathrm{sgn}\,}

\begin{document}
\maketitle
\abstract{We introduce the spinor parallel propagator for maximally
symmetric spaces in any dimension. Then, the Dirac spinor Green's
functions in the maximally symmetric spaces $\mathbb{R}^n$, $S^n$ and
$H^n$ are calculated in terms of intrinsic geometric objects. The
results are covariant and coordinate-independent.}

\newpage
\section{Introduction}
The study of field theory in Anti-de Sitter (AdS) spaces, which
topologically are hyperbolic maximally symmetric spaces, has been
revived over the past two years following the so-called Maldacena
conjecture relating type IIB supergravity on AdS$_5$ $\times$ $S^5$
with $\mathcal{N}=4$, $U(N)$ Super Yang Mills theory in four
dimensions. 

More than a decade ago, the calculation of correlation functions in
maximally symmetric spaces using only intrinsic geometric objects was 
presented in a series of papers starting with
\cite{Burgess85,Allen86a,Allen86b}. In one 
of them \cite{Allen86b}, Green's functions for two-component spinors
in maximally symmetric four-spaces were considered using the $SL(2,R)$
formulation. To our knowledge, this analysis has not been extended
since to Dirac spinors in other space-time dimensions. However, it
should be mentioned 
that spinor Green's functions in AdS spaces have been considered and
calculated by other means in the context of the AdS/CFT correspondence
\cite{Kawano99,Rashkov99}.

In the present paper, we present an intrinsically geometric approach
to spinor Green's functions in maximally symmetric spaces. In
section~\ref{pprop}, we introduce the spinor parallel propagator for
maximally symmetric spaces of dimension $n$ and find its covariant
derivatives. Then, in section~\ref{sprop}, we calculate the spinor
Green's functions for the spaces $\mathbb{R}^n$, $S^n$ and
$H^n$. Finally, section~\ref{conc} contains conclusions.

In the remainder of this section, we would like to review the
elementary maximally symmetric bi-tensors, which
have been discussed in detail by Allen and Jacobsen
\cite{Allen86a}. 

Consider a maximally symmetric space of dimension $n$ with constant
scalar curvature $n(n-1)/R^2$. For the space $S^n$, the radius $R$ is
real and positive, whereas for the hyperbolic space $H^n$, $R=il$ with
$l$ positive, and in the flat case, $\mathbb{R}^n$, $R=\infty$. 

Consider further two points $x$ and $x'$, which can be connected
uniquely by a shortest geodesic. Let $\mu$ be the proper geodesic
distance along this shortest geodesic between $x$ and $x'$. 
Then, the vectors 
\begin{equation}
\label{ndef} 
  n_\nu(x,x') = D_\nu \mu(x,x') \quad \text{and} \quad 
  n_{\nu'}(x,x') = D_{\nu'} \mu(x,x')
\end{equation}
are tangent to the geodesic and have unit length. Furthermore, denote
by $g^\mu_{\;\nu'}(x,x')$ the vector parallel propagator along the
geodesic. Notice the relation $n^{\nu'} = -g^{\nu'}_{\;\mu} n^\mu$. 

These elementary maximally bi-tensors $n^\mu$, $n^{\mu'}$ and
$g^\mu_{\;\nu'}$ satisfy the following properties:
\begin{subequations}
\begin{align}
\label{dn}
 D_\mu n_\nu &= A(g_{\mu\nu} -n_\mu n_\nu)\\
\label{dnprime}
 D_{\mu'} n_\nu &= C(g_{\mu'\nu} +n_{\mu'} n_\nu)\\
\label{dg}
 D_\mu g_{\nu\lambda'} &= -(A+C) (g_{\mu\nu} n_{\lambda'} +
 g_{\mu\lambda'} n_\nu),
\end{align}
\end{subequations}
where $A$ and $C$ are functions of the geodesic distance $\mu$ and are
given by 
\begin{equation}
\label{AC}
 A = \frac1R \cot \frac{\mu}R \quad \text{and} \quad 
 C = -\frac1{R\sin(\mu/R)}. 
\end{equation}
Therefore, they satisfy the relations 
\begin{equation}
\label{ACrel}
 dA/d\mu =-C^2, \quad dC/d\mu =-AC \quad \text{and} \quad C^2-A^2
 =1/R^2.
\end{equation}

Finally, our convention for covariant gamma matrices is
$\{\Gamma^\mu,\Gamma^\nu\} =2 g^{\mu\nu}$.

\section{Spinor Parallel Propagator}
\label{pprop}
To start, consider a bi-spinor $\Lambda(x',x)^{\alpha'}_{\;\beta}$,
which acts as parallel propagator for Dirac spinors in a maximally
symmetric space-time, i.e.\ it performs the parallel transport 
\[{\Psi'}(x')^{\alpha'} = \Lambda(x',x)^{\alpha'}_{\;\beta} \Psi(x)^\beta.\]

The spinor parallel propagator $\Lambda(x',x)$ can be uniquely defined
for any space-time dimension by the following properties:
\begin{subequations}
\begin{align}
\label{lambdadef1}
  \Lambda(x',x) &= [\Lambda(x,x')]^{-1},\\
\label{lambdadef2}
  \Gamma^{\nu'}(x') &= \Lambda(x',x) \Gamma^\mu(x) \Lambda(x,x')
  g^{\nu'}_\mu(x',x),\\
\label{lambdadef3}
  n^\mu D_\mu \Lambda(x,x') &=0.
\end{align}
\end{subequations}
Eqn.\ \eqref{lambdadef1} implies that $\Lambda(x,x)^{\alpha'}_\beta =
\delta^{\alpha'}_\beta$, whereas eqn.\ \eqref{lambdadef2} conveniently
formulates the parallel transport of the covariant gamma
matrices. Finally, eqn.\ \eqref{lambdadef3} says that $\Lambda(x,x')$
is covariantly constant along the geodesic of parallel transport. 

We would like to evaluate now a particular property of
$\Lambda(x,x')$, namely its covariant derivative. Therefore, combine
eqns.\ \eqref{lambdadef1} and \eqref{lambdadef2} to 
\begin{equation}
\label{gammaprop}
 \Gamma^\nu \Lambda(x,x') = \Lambda(x,x') \Gamma^{\mu'} g^\nu_{\mu'}
\end{equation}
and differentiate covariantly with respect to $x$ to obtain
\begin{equation}
\label{defdiff}
 \Gamma^\nu D_\lambda \Lambda(x,x') = D_\lambda \Lambda(x,x') \Gamma^{\mu'}
 g^\nu_{\mu'} - (A+C) \Lambda(x,x') \Gamma^{\mu'} (\delta^\nu_\lambda
 n_{\mu'} + g_{\lambda\mu'} n^\nu), 
\end{equation}
where we have used the property \eqref{dg} of the vector parallel
propagator. Now, use eqn.\ \eqref{gammaprop} for the second term on
the right hand side of eqn.\ \eqref{defdiff} and multiply with
$\Gamma^\lambda$, which yields 
\begin{equation}
\label{defdiff2}
 2 D^\nu \Lambda(x,x') - \Gamma^\nu \slashD \Lambda(x,x') = \slashD
 \Lambda(x,x') \Gamma^{\mu'}g^\nu_{\mu'} + (A+C) (\Gamma^\nu
 \Gamma^\rho n_\rho - n n^\nu) \Lambda(x,x').
\end{equation}
Thus, a multiplication with $\Gamma_\nu$ leads to 
\[ (2-n) \slashD \Lambda(x,x') = \Gamma_\nu \slashD \Lambda(x,x')
\Gamma^{\mu'} g^\nu_{\mu'},\]
the solution of which is
\begin{equation}
\label{slashDlambda}
  \slashD \Lambda(x,x') = B n_\mu \Gamma^\mu \Lambda(x,x'), 
\end{equation}
where $B$ is some function of the geodesic distance $\mu$. 
Then, substituting eqn.\ \eqref{slashDlambda} into eqn.\
\eqref{defdiff2} yields 
\[ 2 D^\nu \Lambda(x,x') = 2 B n^\nu \Lambda(x,x') 
+ (A+C) (\Gamma^\nu \Gamma^\rho n_\rho - n n^\nu) \Lambda(x,x').\]
Moreover, multiplying this with $n_\nu$ and using eqn.\
\eqref{lambdadef3} one determines $B$ to be 
\[ B= \frac12 (n-1)(A+C).\]
Therefore, one finally obtains 
\begin{equation}
\label{dlambda}
  D_\mu \Lambda(x,x') = \frac12 (A+C) \left(\Gamma_\mu \Gamma^\nu n_\nu
  -n_\mu\right) \Lambda(x,x').
\end{equation}

For completeness, we also give the expression for $D_{\mu'}
\Lambda(x,x')$. It is easily obtained from eqn.\ \eqref{dlambda} using
eqn.\ \eqref{lambdadef1} and is given by
\begin{equation}
\label{dlambda2}
  D_{\mu'} \Lambda(x,x') = -\frac12 (A+C) \Lambda(x,x') \left(\Gamma_{\mu'}
  \Gamma^{\nu'} n_{\nu'} - n_{\mu'}\right).
\end{equation}

\section{Spinor Green's Function}
\label{sprop}
Using the spinor parallel propagator $\Lambda(x,x')$ calculated in
section~\ref{pprop}, we would now like to find the spinor Green's
function $S(x,x')$ satisfying 
\begin{equation}
\label{greendef}
  \left[(\slashD -m) S(x,x')\right]^\alpha_{\;\beta'} =
  \frac{\delta(x-x')}{\sqrt{g(x)}} \delta^\alpha_{\beta'}.
\end{equation}
Here, we have written the indices explicitly in order to emphasize
that this is a bi-spinor equation. We shall henceforth omit the
indices. 

Now, we make the general ansatz 
\begin{equation}
\label{greenans}
  S(x,x') = \left[ \alpha(\mu)+\beta(\mu) n_\nu \Gamma^\nu \right]
  \Lambda(x,x'),
\end{equation}
where $\alpha$ and $\beta$ are functions of the geodesic distance
$\mu$ still to be determined. We substitute the ansatz
\eqref{greenans} into eqn.\ \eqref{greendef} and, after using eqn.\
\eqref{dlambda}, obtain the two coupled differential equations 
\begin{align}
\label{sys1}
  \beta' +\frac12(n-1)(A-C) \beta -m\alpha &=
  \frac{\delta(x-x')}{\sqrt{g(x)}},\\
\label{sys2}
  \alpha' +\frac12(n-1)(A+C)\alpha - m\beta &=0,
\end{align}
where the prime denotes differentiation with respect to $\mu$.

In order to proceed, multiply eqn.\ \eqref{sys1}
with $m$ and substitute $m\beta$ from eqn.\ \eqref{sys2}. One finds
\begin{equation}
\label{alphaeq}
  \alpha''+(n-1) A\alpha' -\frac12(n-1)C(A+C) \alpha -\left[
  \frac{(n-1)^2}{4R^2} +m^2 \right] \alpha
  = m \frac{\delta(x-x')}{\sqrt{g(x)}}, 
\end{equation}
where eqn.\ \eqref{ACrel} has been used. We shall solve equation
\eqref{alphaeq} separately for the spaces $\mathbb{R}^n$, $S^n$ and
$H^n$. 

\subsection{Green's Function for \boldmath$\mathbb{R}^n$}
For $\mathbb{R}^n$, we have $A=-C=1/\mu$, $R=\infty$ and
$\mu=|x-x'|$. Thus, eqn.\ \eqref{alphaeq} becomes 
\begin{equation}
\label{alphaeqR}
  \alpha''+\frac{n-1}\mu \alpha' -m^2\alpha
  = m\, \delta(x-x').
\end{equation}
The solution to eqn.\ \eqref{alphaeqR} is 
\begin{equation}
\label{alphaR}
  \alpha(\mu) = - \left(\frac{m}{2\pi}\right)^\frac{n}2
  \mu^{1-\frac{n}2}\, \mathrm{K}_{\frac{n}2-1}(m\mu),
\end{equation}
where the functional form was obtained by solving eqn.\
\eqref{alphaeqR} for $\mu\ne0$, and the constant was found by
matching the singularity. Furthermore, one finds from eqn.\
\eqref{sys2} $m\beta =\alpha'$, i.e.\ $n_\nu \beta = \partial_\nu
\alpha/m$, so that the final result for the spinor Green's function
in $\mathbb{R}^n$ is 
\begin{equation}
\label{greenR}
  S(x,x') = - \frac1m \left(\frac{m}{2\pi}\right)^\frac{n}2
  (\slashdel + m) \mu^{1-\frac{n}2}\,
  \mathrm{K}_{\frac{n}2-1}(m\mu). 
\end{equation}
Upon Fourier transforming it, one obtains the more familiar expression 
\begin{equation}
\label{greenRfour}
  S(x,x') = - (\slashdel + m)
  \int\frac{d^n\!k}{(2\pi)^n}\; \mathrm{e}^{-i
  k\cdot(x-x')} \frac1{k^2+m^2}.
\end{equation}

\subsection{Green's Function for \boldmath$S^n$}
In order to solve eqn.\ \eqref{alphaeq}, we consider first $x\ne x'$
and make the substitution  
\begin{equation}
\label{zsub}
  z = \cos^2 \frac{\mu}{2R}.
\end{equation}
This yields the differential equation
\begin{equation}
\label{alphaeqS}
  \left[ z(1-z) \frac{d^2}{dz^2} + \frac{n}2 (1-2z)\frac{d}{dz} -
  \frac{(n-1)^2}4 -m^2 R^2 - \frac{n-1}{4z} \right] \alpha(z) =0.
\end{equation}
Then, writing $\alpha(z)=\sqrt{z}\gamma(z)$, one obtains a
hypergeometric equation for $\gamma$,
\begin{subequations}
\label{gammaeq}
\begin{gather}
\label{gammaeq1}
  H(a,b;c;z) \gamma(z) = 0,\\
\intertext{where}
\label{H}
  H(a,b;c;z) = z(1-z) \frac{d^2}{dz^2} + [c-(a+b+1)z] \frac{d}{dz}
  -ab\\
\intertext{is the hypergeometric operator, and its parameters are} 
\label{abc}
  a = \frac{n}2 -i|m|R, \quad b = \frac{n}2+i|m|R, \quad c =
  \frac{n}2+1.
\end{gather}
\end{subequations}
The solution of eqn.\ \eqref{gammaeq} which is singular at $z=1$ 
is \cite{Gradshteyn} 
\begin{equation}
\label{gammaS}
  \gamma(z) = \lambda \,\mathrm{F}(a,b;c;z) = \lambda
  \,\mathrm{F}(n/2-i|m|R,n/2+i|m|R;n/2+1;z),  
\end{equation}
where $\lambda$ is a proportionality constant. Therefore, $\alpha(z)$
is 
\begin{equation}
\label{alphaS}
  \alpha(z) = \lambda \sqrt{z}\,\mathrm{F}(n/2-i|m|R,n/2+i|m|R;n/2+1;z).
\end{equation}
We can now determine the constants $\lambda$ by matching
the singularity in eqn.\ \eqref{alphaeq}. This is equivalent to
demanding the singularity of $\alpha$ at $\mu=0$ to have the same
strength as in the case of $\mathbb{R}^n$. One finds from eqn.\
\eqref{alphaS} 
\[ \alpha \to \lambda
\frac{\Gamma(n/2+1)\Gamma(n/2-1)}{\Gamma(n/2-i|m|R)\Gamma(n/2+i|m|R)} 
\left(\frac{\mu}{2R}\right)^{2-n},\]
whereas in $\mathbb{R}^n$ we have, from eqn.\ \eqref{alphaR},
\begin{equation}
\label{Rasym}
  \alpha \to - \frac{m}4 \Gamma(n/2-1) \pi^{-n/2} \mu^{2-n}.
\end{equation}
Comparing these two expressions we find
\begin{equation}
  \lambda = - m
  \frac{\Gamma(n/2-i|m|R)\Gamma(n/2+i|m|R)}{\Gamma(n/2+1) \pi^{n/2}
  2^n} R^{2-n}.
\end{equation}
Finally, one can calculate $\beta$ from eqn.\ \eqref{sys2}, which yields
\begin{align}
\notag
  \beta(z) &= -\frac1m \left[ \frac1R \sqrt{z(1-z)} \frac{d}{dz} +
  \frac{n-1}{2R} \sqrt{\frac{1-z}z} \right] \alpha(z)\\
\label{betaS}
  &= -\frac{\lambda}{mR} \sqrt{1-z} \left[ z \,
  \mathrm{F}(n/2+1-i|m|R,n/2+1+i|m|R;n/2+2;z) \phantom{\frac{n}2}
  \right. \\ \notag &\quad 
  + \left.\frac{n}2 \, \mathrm{F}(n/2-i|m|R,n/2+i|m|R;n/2+1;z) \right].
\end{align}
It should be noticed that $\beta$ has a finite $m\to0$ limit, whereas
$\alpha$ vanishes. 

\subsection{Green's Function for \boldmath$H^n$}
For $H^n$, we can start with eqn.\ \eqref{gammaeq} and set $R=il$,
i.e.\ we have to solve
\begin{subequations}
\label{gammaeqH}
\begin{gather}
\label{gammaeqH1}
  H(a,b;c;z) \gamma(z) = 0\\  
\intertext{with} 
  a = \frac{n}2 + |m|l, \quad b = \frac{n}2-|m|l, \quad c =
  \frac{n}2+1. 
\end{gather}
\end{subequations}
There are two solutions to eqn.\ \eqref{gammaeqH} which behave 
asymptotically like a power of $z$ for $z\to\infty$. These are
\begin{equation}
\label{gammaH}
  \gamma_\pm(z) = \lambda_\pm z^{-\left(\frac{n}2 \pm |m|l\right)}
  \,\mathrm{F} \left(\frac{n}2\pm |m|l, \pm |m|l; 1\pm 2|m|l; \frac1z
  \right),
\end{equation}
where $\lambda_\pm$ are constants. The choice of the minus sign is not
always possible. In fact, for $1-2|m|l=0,-1,-2,\ldots$ the
hypergeometric series is indeterminate. Thus, we shall include the
solution with the minus sign only, if $|m|l<1/2$. Hence, we have
two solutions for $\alpha$,
\begin{equation}
\label{alphaH}
  \alpha_\pm(z) = \lambda_\pm z^{-\left(\frac{n-1}2 \pm |m|l\right)}
  \,\mathrm{F} \left(\frac{n}2\pm |m|l, \pm |m|l; 1\pm 2|m|l; \frac1z
  \right),
\end{equation}
and we can now proceed to determine the constants $\lambda_\pm$ in a
similar fashion as in the $S^n$ case. 
From eqn.\ \eqref{alphaH} we find for $\mu\to0$
 \[ \alpha \to \lambda_\pm \left(\frac\mu{2l}\right)^{2-n}
 \frac{\Gamma(1\pm2|m|l) \Gamma(n/2-1)}{\Gamma(n/2\pm|m|l)
 \Gamma(\pm|m|l)}.\]
Comparing this expression to the $\mathbb{R}^n$ case, eqn.\
\eqref{Rasym}, we find
\begin{equation}
\label{lpm}
  \lambda_\pm = \mp \sgn m\, 2^{-(n\pm2|m|l)}
  l^{1-n}\frac{\Gamma(n/2\pm|m|l)}{\pi^{(n-1)/2} \Gamma(1/2\pm|m|l)},
\end{equation}
where the doubling formula for Gamma functions has been used.

Finally, let us calculate $\beta$ from eqn.\ \eqref{sys2}. Using a
recursion formula for hypergeometric functions we find
\begin{align}
\notag
  \beta_\pm(z) &= \frac1m \left[ \frac1l \sqrt{z(z-1)} \frac{d}{dz} +
  \frac{n-1}{2l} \sqrt{\frac{z-1}z} \right] \alpha_\pm (z) \\
\label{betaH}
  &= \mp \sgn m\, \lambda_\pm \sqrt{z-1}\, z^{-\left(\frac{n}2 \pm
  |m|l\right)} \,\mathrm{F} \left(\frac{n}2\pm |m|l, 1\pm |m|l; 1\pm
  2|m|l; \frac1z \right).
\end{align}
It is interesting to note that in the limit $m\to0$ the functions
$\beta_+$ and $\beta_-$ become identical, whereas $\alpha_+$ and
$\alpha_-$ do not, but differ in their signs. The reason is, of
course, that, for $m=0$, eqns.\ \eqref{sys1} and \eqref{sys2}
decouple, and $\alpha$ can be a solution of eqn.\ \eqref{sys2} with
arbitrary proportionality constant. 
Moreover, for $m=0$, the common value of $\beta_\pm$ is a rational
function of $z$,
 \[ \beta_\pm (z)= \frac{\Gamma(n/2)}{(2\pi)^n} l^{1-n}
 (z-1)^{-(n-1)/2}.\]

\section{Conclusions}
\label{conc}
We have introduced the spinor parallel propagator for maximally
symmetric spaces in any dimension. This enabled us to 
find expressions for the Dirac spinor Green's functions in the
maximally symmetric spaces $\mathbb{R}^n$, $S^n$ and $H^n$ in terms
of intrinsic geometric objects. Although there are obstructions to the
quantization of spinors in odd dimensional manifolds with boundary
\cite{Carey99}, our results should be applicable to the AdS/CFT
correspondence, because of the classical dynamics in AdS space.

\section*{Acknowledgments}
I would like to thank K.~S.~Viswanathan and R.~C.~Rashkov for
stimulating discussions. Moreover, financial support from Simon Fraser
University is gratefully acknowledged.

\end{document}